# Exploring Opportunities in Usable Hazard Analysis Processes for AI Engineering

**Nikolas Martelaro,[1] Carol J. Smith,[2] Tamara Zilovic[1]**

HCI Institute - Carnegie Mellon University,[1] Software Engineering Institute, Carnegie Mellon University[2]

nikmart@cmu.edu, cjsmith@sei.cmu.edu, tzilovic@andrew.cmu.edu

**Abstract**

Embedding artificial intelligence into systems introduces significant challenges to modern engineering practices. Hazard analysis tools and processes have not yet been adequately adapted to the new paradigm. This paper describes initial research and findings regarding current practices in AI-related hazard analysis and on the tools used to conduct this work. Our goal with this initial research is to better understand the needs of practitioners and the emerging challenges of considering hazards and risks for AI-enabled products and services. Our primary research question is: *Can we develop new structured thinking methods and systems engineering tools to support effective and engaging ways for preemptively considering failure modes in AI systems?* The preliminary findings from our review of the literature and interviews with practitioners highlight various challenges around integrating hazard analysis into modern AI development processes and suggest opportunities for exploration of usable, human-centered hazard analysis tools.

## Introduction

Products and services are increasingly using artificial intelligence (AI), and specifically machine learning (ML), to enable automation. While the use of AI-based systems can spur innovation, the combination of data and algorithms that underlie complex systems can lead to failures. Prominent examples include computer vision algorithms that label Black people as gorillas (Barr, 2015; Mac, 2021), chatbots that speak out hate (Lee, 2016), and autonomous vehicle vision systems whose inability to distinguish trucks from the sky has led to fatal crashes (Yadron & Tynan, 2016). Such failures in AI systems not only pose significant risks to the end users of such products but can have profound financial and reputational implications for the organizations that develop these products. While the use of new technologies always comes with the possibility of unintended consequences, we believe that many of these examples could have been prevented through strategic and thoughtful consideration when these systems are being designed and engineered.

Within systems engineering, strategies for hazard analysis can be used by teams to identify risks and potential failures with the goal of developing more robust and safe engineered systems. While many formal hazard analysis techniques exist, these activities largely center around helping teams determine potential risks and/or sources of failure *before* products have begun the development process. However, Rae at al. (2020) have called into question the utility of such methods as it is unclear if and how such methods are used by practitioners (Provan et al., 2017).

Within the space of AI-enabled systems, there has been recent work to support practitioners in developing more ethical practices and fairer ML models (Holstein et al 2019; Madaio et al 2020) and to consider safety in machine learning system development in ways similar to other industries (Varshney, 2016; Varshney & Alemzadeh, 2017). While such work is spurring the development of new tools and techniques to support developing fairer ML-systems (e.g., Datasheets for Datasets (Gebru et al., 2021) & Model Cards for Model Reporting (Mitchell et al., 2019)), these methods are focused on the early stages of development and thus may be less integrated into the entire engineering process for AI-enabled products and services.

In this paper, we describe our initial findings from a set of exploratory interviews that suggest that although teams do generally recognize the value of hazard analysis - and do try to employ practices aimed at early recognition and communication of potential risks - various shared challenges regarding the development and implementation of these processes have been reported. Common themes that ran through our interviews included the incompatibility of such processes with modern development practices, unique challenges posed by working with non-deterministic ML systems, limited tooling available to support such activities, time pressures inherent to competitive markets, and the role of company culture in the support of these efforts.

Based on these initial findings, we believe that there are exciting opportunities to adapt and extend more traditional hazard analysis processes for use by fast-moving AI engineering teams, and to develop new hazard analysis tools and processes tailored towards addressing the emergent needs of developing AI-enabled products. Motivated by Rae et al. (Rae et al., 2020) and their call for reality-based safety research, we look to understand the needs of AI engineering teams and aim to develop new, usable hazard analysis tools. We are currently in the initial stages of this research.

## Related Work

### Hazard analysis techniques and their use

According to prior work cited by Rae at al. (2020), it is suggested that most professional hazard analysis practice is driven by regulation (Provan et al., 2019; Van Eerd et al., 2018). As such, hazard analysis activities are more common in regulated industries such as aerospace, automotive, finance, and healthcare. Various techniques for hazard analysis such as Failure Modes and Effects Analysis (FMEA), Fault Tree Analysis (FTA), Hazard and Operability Analysis (HAZOP), and System Theoretic Process Analysis (STPA) are common.

However, the lack of clarity regarding how practitioners use such techniques and the seeming disconnect with professional practice have led Rae at al. (2020) to call into question the utility of some hazard analysis methods. Provan et al. (Provan et al., 2017) suggest that there has been little work researching professional practice and as Rae et al. point out, professional hazard analysis practice appears to be driven by necessity to comply with existing regulation rather than by new research that focuses on the utility of hazard analysis theory (Provan et al., 2017; Van Eerd et al., 2018). Such work suggests a need to better understand how professionals consider hazards and risks throughout their design process and to develop tools that support such work effectively. As part of our research, we look to understand how AI engineering teams may or may not incorporate hazard analysis processes into their work currently.

### Responsible AI and new techniques for considering risks

Organizations and researchers working on AI systems have been exploring new ways to engage practitioners in thinking about responsible AI. A common starting point is adopting an existing ethical framework from a reputable organization focused on computing and AI (Smith, 2019). The Montréal Declaration for a Responsible Development of Artificial Intelligence (Fjeld et al., 2020; Hagendorff, 2020; Université de Montréal, 2018), is a respected starting place. Ethical principles designed for AI use can help to bridge gaps between team members' personal experiences and may be particularly helpful in supporting cross-functional and diverse teams. Two efforts to compare sets of AI ethics (Fjeld et al., 2020; Hagendorff, 2020) found broad patterns across the principles, some of which dealt with the success of professional implementation. Hagendorff (2020) identified issues in implementation of AI ethics due to the lack of technological detail, a lack of reinforcement mechanisms, and an insignificant influence on the decision-making of individual software developers.

Adopting AI ethics is not enough - targeted guidance, tools and processes must be provided to the team to support the process of evaluating risk (Dunnmon et al., 2021; Smith, 2019). Such a holistic approach can facilitate conversations regarding the scope of the system's capabilities, support the team in speculating about harmful consequences, assist in the creation of plans to prevent potential harms, and/or establish mitigation plans to stop unforeseen harm quickly. Though this work is difficult for teams to engage in, it does not have to be onerous.

To facilitate conversations around harms, various new tools and techniques are emerging that are aimed at engaging teams in considering potential hazards while using AI-based technologies. A key aspect in reducing harms from ML systems is establishing a thorough understanding of the data selected for the system. A set of data that is either inappropriate to the function of the system–or whose breadth is too narrow or too broad–can create unsolvable problems if not identified (and remedied) early in the process. The Datasheets for Datasets method proposed in 2018 by Gebru et al. (2021) guides a team in identifying the data creator, their motivation for the data collection, and what data is included or excluded - among other important details - so that teams can consider if the dataset is appropriate for their task. The same individuals then continued the effort by developing Model Cards for Model Reporting that include clear descriptions of the intent and functionality of the model(s) an AI system is based on (Mitchell et al., 2019). Such methods can aid AI engineers in using readily available data and models more thoughtfully.

As engineers design and develop new models, many responsible AI processes suggest considering harms early in the process. For example, in the Microsoft AI Fairness Checklist (Microsoft Research, 2021), Madaio et al. (2020) highlight envisioning the types of harms and the potential tradeoffs between benefits and harms as part of the first major phase of new AI-driven projects. While no explicit methods are given in the checklist on methods for envisioning harms, other research has looked at creating techniques for fostering conversation. One notable area in this space are card games designed to help provide structure around thinking about harms while doing so in a less formal and more engaging way. Such examples include Ballard et al.'s (2019) "Judgement Call", which uses value sensitive design to help surface ethical issues for ML systems and Martelaro and Ju's (2020) "What Could Go Wrong?" which aims to help teams in considering the possible effects and failure modes in autonomous vehicles and services enabled by autonomous vehicles. Overall, such methods aim to engage teams in thinking about potential harms and what faults may be associated with them. Conversations facilitated by such games can be the start of more formal hazard analysis processes and engage more diverse stakeholders in the process of considering risks for AI systems. Inspired by this prior work, our research looks to extend thinking around new methods and processes to enable teams to better consider and design for potential AI-systems hazards.

## Motivation

We see preventable examples of system failures frequently in the media. These failures are a risk to the users of such products and can have financial and reputational implications for the organizations that develop them. Many of these examples could be preventable with strategic and thoughtful consideration at the time of design. While there are structured systems engineering methods to support thinking about failures, these are not always conducted due to lack of experience, interest, and/or perceived lack of time. We intend to understand where there are shortcomings in current professional practice and then to develop tools aimed at supporting critical thinking around possible hazards in AI engineering. Our primary research question is: *Can we develop new structured thinking methods and systems engineering tools to support effective and engaging ways for preemptively considering failure modes in AI systems?*

This work is part of the Carnegie Mellon University Software Engineering Institute's effort leading a national AI engineering initiative to define a professional discipline for AI engineering. As human-computer interaction (HCI) practitioners and academics, we look to engage people in relatable ways of thinking about issues and failures that can arise during AI product development. Based on our own work practices and the call for reality-based safety science by Rae et al. (2020), we approach our work through a human-centered design perspective. This includes defining a specific user, clearly defining the needs and tasks of that user that the system will support and outlining a set of baseline metrics within which to base the system's expected performance (Dunnmon et al., 2021).

*Outline of our research goals and methods*

We are conducting this work in four phases over the course of a year. Our first phase includes background research on the latest hazard analysis techniques and a set of exploratory interviews with practitioners developing complex engineering systems to understand how such methods are used in practice. Our second phase of work will involve gathering additional information regarding hazard analysis via a survey and designing and prototyping useful and usable new tools and interfaces to enable fruitful conversations and tracking of potential harms by AI engineering teams. We will then test our prototypes with practitioners through a series of iterative testing and feedback sessions. Finally, we will test our designs through a longer-term engagement with practitioners. Currently, we are ending our first phase of work and report on our initial literature review and exploratory interviews.

## Exploratory Interviews

### Method

We have completed a set of eight semi-structured interviews with practitioners and instructors whose work has some relation to considering the potential harms from AI-based systems. Interviews were 30-minutes long and centered on the processes that our interviewees typically take in thinking about hazard analysis and risk, how current hazard analysis processes work or do not work well, how considering hazard analyses for AI-based systems may be different from other complex systems, and what challenges and frustrations professionals have with current hazard consideration practices (see Appendix A: Interview Guide).

We recruited participants who worked in areas dealing with complex engineered systems based on personal connections and referrals. We emailed potential participants and allowed them to schedule a time using an online calendaring system. All but one of our interviews were conducted over video conferencing. All participants signed a consent form approved by CMU IRB STUDY2021_00000274. Interviews were recorded unless participants did not consent to record. A research team member took written notes during the interview. After the interview, we extracted insightful and interesting quotes and comments to support our qualitative analysis.

### Participants

At the present time, we have spoken to eight people who work in the area of engineering complex systems. Four are in industry, three consult to industry clients, and one was an instructor teaching AI software engineering. Our participants represented the following roles: co-founder of a startup, safety lead at a large autonomous vehicle company, senior research scientist at a large software company, engineering manager at a software company, principle of ethics and innovation at a top-five consulting company, consultant for a government funded research organization, senior engineer at a policy think tank, and the professor at a R1 university.

We note that, so far, discussing practitioners' processes around considering hazards has been challenging. Our eight participants have come from reaching out to a pool of 37 people - while generally a good response rate, we expected a higher rate due to our connections with them. Beyond the challenges of scheduling interviews with professionals, the subject matter of our interview is likely sensitive as it concerns people discussing their knowledge of risks which can bring about liability. For example, two participants requested to see the questions beforehand to ensure approval to speak about the area. Future research should consider the sensitivity of this subject matter when recruiting professionals.

**Findings**

Our initial interviews–paired with our background research–have highlighted various challenges and shortcomings in current practices for considering potential risks and harms. Below, we outline our primary findings.

*Lack of incentives*

Our interviews revealed that formal processes are primarily used by teams operating within regulated industries that need to comply with formal standards, such as aerospace and finance. Non-regulated industries, and organizations creating technology that is not obviously part of a physically risky system, are often not incentivized to do this work. The perception of hazard analysis as being very time-consuming–combined with ambiguous understandings of best practices–often leads to this work being omitted. Additionally, organizations have been discouraged from looking at hazards to avoid culpability, creating a further disincentive.

*Lack of cross-functional teams*

Some of our interview participants identified engineers as being the primary people involved in hazard analysis work. While some indicated the presence of defined safety teams within their organizations, others mentioned the expectation that safety analysis was to be distributed among all employees. Regarding the communication of potential safety concerns, it was noted that some companies do seem to have formal processes for employees to report potential hazards, such as submitting white papers on potential risks. Interestingly, one participant noted that at a company within which they had worked, there were financial incentives attached to participating in this process; if a reported potential failure did occur, that employee was given a bonus.

Cross-disciplinary participation in safety analysis was not evident in practice among the collective experiences of our interviewees. One participant - who initially indicated the expectation of safety considerations to be distributed among all employees - noted that a designer on their team was typically excluded from hazard analysis conversations *"because he doesn't have AI in his title."*

*Culture as a driver of action*

Our interviewees reported that culture can be a positive influence for team members considering risks - making safety thinking a part of everyone's job, having incentives for safety thinking (bonuses distributed for submitting documents describing hazards that are later identified), and by prioritizing metrics that go beyond sales.

However, sometimes there are aspects of culture that have a negative impact on this work. Not all of the information relevant to considering harms from engineered AI systems may be shared with relevant team members - one systems engineer participant discussed frustrations around how other engineers might not share all the details with a safety team due to a perception that the system engineering team was not "technical enough." Additionally, while hazard analyses may be conducted, they may not have the intended positive impacts. One of our consultant participants complained that organizations, in some cases, are conducting hazard analysis work primarily to placate newer talent rather than as a sustainable and systemic long-term strategy for risk management during product development.

*Rudimentary tools*

Among our participants, we did not hear about the use of specialized software tools for conducting hazard analyses or supporting hazard analysis thinking. While some participants did mention named hazard analysis processes, teams seemed to use rudimentary tools (spreadsheets) to document and track work. While functional, one interviewee said that such tools limit their ability to conduct high fidelity tracing of their efforts aligned with the development of the system. Another participant also suggested that tools for documenting system processes and hazards did not have a particularly effective way for visualizing how humans may interact with and develop a mental model of the system over time. The challenge of mode confusion during system use was raised as a particularly arduous interaction to document and consider during a hazard analysis. Furthermore, tracking hazards can be challenging in systems that require consistent manual updating, like spreadsheets. Interviewees noted the importance of - but lack of support for - traceability within their current tooling.

*Working at the speed of development*
One of our participants - who developed a new hazard analysis method and attempted to employ it at a company - stated that the company appreciated the thinking it supported, but found the method was too complicated to be done efficiently within the team's timeline. Other interviewees suggested that the speed and iterative nature of modern software development does not support the meticulous processes that existing hazard analysis frameworks entail. One participant said that although safety analysis would ideally be completed up front, this is not always possible when the technological capabilities are still being developed. They commented that the work is never truly done - as soon as you finish reviewing the existing work, you must start again. Autonomous systems and AI/machine learning systems - especially those built-in ways that are not easily scrutinized - are particularly challenging. Team members conducting hazard analyses may not understand the inner workings of the algorithms and have difficulty breaking the system into components to understand the source of the issue.

*Underestimating complexity*
One participant has observed that smaller organizations underestimate the engineering needed to integrate AI into production and noted that much thinking about hazards focused solely on individual models without considering how these models were part of a larger product system. It was noted that many small companies only have data scientists and expect them to also manage the AI engineering responsibilities. However, these data scientists may feel that taking on the burden of engineering large software systems falls out of the scope of their responsibilities (or interests) or they may not have experience developing large software systems. Inadequate engineering of the AI system may result in aspects being less robust than required for real-world production use.

Another interviewee discussed the failure of people doing the work to think at a systems level. They discussed how minor changes can have broad implications and a lack of exponential thinking about potential consequences can be an issue. They also discussed the overall challenges of having conversations around risk and scale - and the impact of changes that can affect millions of people.

Finally, multiple interviewees mentioned that working with ML-based systems was different from deterministic control systems. Many ML models are inherently probabilistic and are thus challenging to envision and test for all possible edge cases. *"You can't give [the model] all the examples that are going to happen in the world… If you could, then you would be God"* stated one participant. This participant also discussed the difficulty of managing end-user expectations; even if systems are designed to be safe within limits, end-users may overestimate the AI's capability. Modeling user overconfidence and understanding how design should respond to this is particularly challenging. It was mentioned that there should be better ways of educating users on the limits of an AI-based system.

*Limited investment*
We observed a pattern across the interviewees' responses that suggested limited budgets allotted in some organizations for the purposes of doing hazard analysis work. One participant - who was directly involved in their company's formal hazard analysis process - mentioned that although their company was aware of existing specialized software that could support their efforts, they were not willing to meet the cost required to acquire it. This participant noted that their team thus relied on spreadsheets to track their progress, and also mentioned that the existing specialty software on the market seemed a bit too "clunky."

## Discussion

Through our background research and interviews, we have found that of the people we spoke with, many of Rae et al.'s (2020) critiques on the limited use of formal hazard analysis processes were corroborated. Of our interviewees, only one mentioned a formal and tracked process used within the company. However, they mentioned that it was still challenging to implement and keep up with the company's iterative development processes. One of the consultant participants discussed how a client faced challenges training and providing support to personnel around STPA processes. Moreover, while the software engineering instructor we spoke to did teach hazard analysis techniques (specifically FTA) in class, they did not see evidence that students were taking such lessons into their future work.

Still, on a positive note, we did find that all our participants considered thinking about risks and hazards to be important and did discuss some ways of attempting to address the safety of their products and services. While our participant pool was likely inclined to be thinking about how to consider hazards, we still believe that their motivations do represent other practitioners and their desire to create safe AI-based systems. Our research suggests that the limited use of formal processes stems primarily from the inability of existing processes/frameworks to meet the needs of today's practitioners, whose desire to conduct robust hazard analysis indicates an urgent need to develop modern tools/practices to support this work.

Regarding tools, we have found that prior work in responsible AI focuses on data and model development and the fairness and biases of AI models. While this is an essential component of developing safe AI systems, our participants noted that it is only one part of a larger product system. There is a need to support practitioners in thinking more broadly about entire product systems and the people who are using these systems. Future tools and processes should support engineers in thinking about how end-users may think about and behave with AI-based systems. The authors have experienced first-hand the development of AI systems without consideration of the end-users' experience until late in development. This is a bad practice and additionally, the end-users who are susceptible to autonomy bias, may not be aware of (or understand) the capabilities and limitations of the technology due to its poor design. This can potentially lead to preventable misuse or abuse of the system. We see an opportunity for user experience professionals to integrate hazard analysis efforts into organizations and Responsible AI efforts, as they are closest to the end-users who may experience the hazard first-hand.

Many of our interviewees stressed the role that changing company culture may play in the development of safer and more responsible technology. One participant noted *"you can't solve AI with AI,"* before suggesting that developing safer systems required safety-oriented processes (and likely more robust regulation). From our perspective as HCI professionals and academics, we believe that new tools and processes can help to shape culture within an organization. For example, reducing the formality and engaging multiple stakeholders in some aspects of hazard analysis through lightweight activities such as card games (Ballard et al., 2019; Martelaro & Ju, 2020) could help to shift people's thinking and motivate more formal processes. Additionally, tooling that better integrates into modern development practices and supports continual review and monitoring may also shift team members' thinking about hazard analysis towards something that happens throughout the product development lifecycle and keeps such thinking front of mind.

## Future Work

Based on our initial findings, we are now looking towards confirming what we have found and developing a series of prototypes for new tools and processes to support AI engineering teams in thinking about hazard analysis. We are currently developing a survey to send to AI engineering professionals to learn about the processes that they use and to confirm if the specific challenges that we have seen in carrying out hazard analyses are consistent across industries. We are also conducting a survey of course materials in Computer Science to understand at what points hazard analyses may be introduced to students.

As our research goal is to develop new tools, we will also design and prototype new hazard analysis tools with a focus on two key areas: 1) addressing the formality versus the ease of creating conversations about risk among many stakeholders and 2) developing human-centered interfaces that fit into modern software development practices. Our goals in prototyping new systems will be to further learn about the challenges of doing hazard analyses for AI systems and to see how tools can change practice for the better.

## Conclusion

As AI-based systems become more common in products and services, it will be important for teams to manage associated risks and design to reduce hazards. While existing hazard analysis practices can inform this work, the challenges of modern software development - paired with the sensitive nature of probabilistic AI systems - reveal the need to create new tools and processes that better align with real-world AI engineering practice. As HCI researchers, we believe that useful and usable hazard analysis tools can help teams consider potential hazards and shift their culture towards continual safety consideration.

# Appendix A - Interview Guide

1. Let's start by having you introduce yourself and telling me about your role.
   [If it doesn't come up]
   a. How does your work relate to hazard analysis or risk assessment?
   b. How do you define hazard analysis?

2. Tell me how hazard analysis is used in your organization.
   [If it doesn't come up]
   a. Can walk me through what a typical hazard analysis activity/assessment looks like on your team?
   b. What type of individuals (roles/responsibilities) are typically involved in the hazard analysis work?
   c. Who else should be involved in your opinion? (Who is missing?)
   d. How long has the group been doing aspects of hazard analysis?

3. What methods have you successfully used for hazard analysis? Why were they successful?

4. In your opinion, what is different about hazard analysis for AI systems?

5. What do you think is currently missing from conversations about the implementation of hazard analysis for AI systems?

6. What are your biggest frustrations with hazard analysis?
   a. What is the greatest need for improvement of hazard analysis practices?

*End Questions (skip to Section 4 if time allows)*

7. Who else should we talk to in your organization or elsewhere?

8. What else should I have asked you related to AI engineering?

*Closing*
Thank you very much for your time and responses. I really appreciate it!

*If time allows – then back to End Questions*

9. How does your organization share your hazard analysis work with your customers or stakeholders?

10. What has been your biggest success in AI engineering?

11. What are your biggest AI engineering challenges now?

12. What are your concerns for the future of AI engineering?

13. Have you thought about integrating AI engineering with the rest of the organization?

14. How much is the organization investing in AI engineering (e.g. more money, staff, etc.) vs the total investment in AI (looking for percentages)